\documentclass[aip,amsmath,amssymb,reprint]{revtex4-1}

\usepackage{graphicx}
\usepackage{dcolumn}
\usepackage{bm}
\usepackage[utf8]{inputenc}
\usepackage[T1]{fontenc}
\usepackage{mathptmx}
\usepackage{mathbbol}
\usepackage{xspace}
\newcommand{\Dt}{k}
\newcommand\subcap[1]{(#1)}

\newcommand\subfig[2]{{Fig.~\ref{#1}{(#2)}}}
\newcommand\secti[1]{{Section~\ref{#1}}}
\renewcommand\vec[1]{\bm{#1}}

\newcommand\app[1]{Appendix~\ref{#1}}
\newcommand{\eq}[1]{eq.~\eqref{#1}}
\newcommand{\eqtwo}[2]{eqs~\eqref{#1} and~\eqref{#2}}

\newcommand{\fig}[1]{Fig.~\ref{#1}}

\newcommand{\quot}[1]{``#1''}
\newcommand{\tab}[1]{Table~\ref{#1}}

\newcommand{\expb}[1]{\exp \glb #1 \grb} 
\newcommand{\glb}{\left(}  
\newcommand{\grb}{\right)}  

\newcommand{\VEC}[1]{\mathbf{#1}}
\newcommand{\rvec}{\VEC{r}}

\newcommand{\xvec}{\VEC{x}}

\newcommand{\lambdatilde}{\tilde{\lambda}}

\newcommand{\epsvec}{\boldsymbol{\varepsilon}}

\usepackage{xcolor}

\begin{document}

\title[A kinetic-Monte Carlo perspective on active matter]{A kinetic-Monte 
Carlo perspective on active matter}

\author{Juliane U. Klamser}
 \email{Juliane_klamser@yahoo.de.}
 \affiliation{Laboratoire de Physique Statistique, Département de physique de 
l'ENS, 
Ecole Normale Supérieure, PSL Research University, Université Paris Diderot, 
Sorbonne Paris Cité, Sorbonne Universités, UPMC Univ. Paris 06, CNRS, 75005 
Paris, France.}
\author{Sebastian C. Kapfer}%
 \email{Sebastian.kapfer@fau.de.}
\affiliation{Theoretische Physik 1, FAU Erlangen-Nürnberg, Staudtstr. 7, 91058 
Erlangen, Germany}

\author{Werner Krauth}
\email{Werner.krauth@lps.ens.fr.}
\affiliation{Laboratoire de Physique Statistique, Département de physique de 
l'ENS, 
Ecole Normale Supérieure, PSL Research University, Université Paris Diderot, 
Sorbonne Paris Cité, Sorbonne Universités, UPMC Univ. Paris 06, CNRS, 75005 
Paris, France.}

\date{\today}

\begin{abstract}
We study non-equilibrium phases for interacting two-dimensional 
self-propelled particles with 
isotropic pair-wise interactions using a persistent kinetic Monte Carlo (MC)
approach. We establish the quantitative phase 
diagram, including the motility-induced phase separation (MIPS) that is a 
commonly observed collective phenomena in active matter. In addition, we 
demonstrate for several different potential forms
the presence of two-step melting, with an 
intermediate hexatic phase, in regions far from equilibrium. 
Increased activity can melt a two-dimensional solid and the melting 
lines remain disjoint from MIPS. We establish this phase diagram for 
a range  of the inter-particle potential stiffnesses, and  identify
the MIPS phase even in the hard-disk limit. We establish that the full
description of the phase behavior requires three independent control parameters.
\end{abstract}

\maketitle

\section{Introduction\label{sec:Introduction}}

Active matter is an important field of research that considers
particle systems whose microscopic components are characterized by systematic 
persistent dynamic rules and by various types of mutual interactions. 
On a microscopic scale, the persistent dynamics 
breaks the detailed-balance condition (underlying all of 
equilibrium physics) and  defines active matter
as out-of-equilibrium systems.

Many different models of active matter have been proposed. They feature a 
wide range 
of self-propelled dynamics and of mutual interactions. A great many
theoretical studies employ either Langevin or molecular 
dynamics\cite{CatesReview2012,KineticModel,FilyMarchetti2012,CahnHilliard,ActiveCrystalizationLoewen2012,Dumbbell1} 
in order to model persistent motion. Recently, a kinetic Monte Carlo (MC) 
approach was proposed\cite{BerthierHardSpheres,KlKaKr2018} as a minimal model 
for active matter in two dimensions.

Although the primary interest in theoretical models of active matter comes 
from the non-equilibrium 
nature, their properties can often be connected to their equilibrium 
counterparts that are realized in the zero-persistence limit.
This limit is of particular interest in two dimensions, where 
equilibrium
particle systems with short-range interactions cannot crystallize\cite{Mermin}. 
Nevertheless, it 
was established that, from the high-temperature (low-density) liquid regime 
towards the low-temperature (high-density) solid regime, two-dimensional 
equilibrium particle systems normally undergo two phase 
transitions\cite{HalperinNelson1978,NelsonHalperin1979,Young1979}. These 
transitions describe the passage into and out of a hexatic phase that is 
sandwiched between the liquid and the solid phase (see 
\tab{tab:PhaseCharacteristics}). In this work, we are concerned with 
two-dimensional models with repulsive inverse-power-law interactions, for which 
the two-step melting 
scenario in equilibrium is firmly 
established\cite{BernardKrauth2011,KapferKrauth2015}. 

In this work, we extend our earlier findings for a special 
case\cite{KlKaKr2018} 
and show that kinetic MC generically reproduces 
motility-induced phase separation (MIPS) in two-dimensional active-particle 
systems with a wide range of inverse-power-law interactions including the 
hard-sphere limit. We also confirm the stability of the hexatic phase up to 
considerable values of the activity, and conjecture that it is indeed stable at 
any value of the activity. We finally confirm that 
MIPS is generically a liquid--gas transition under the kinetic MC dynamics. 
Moreover, it is decoupled from the 
melting transitions. This separation can be understood in the limit of 
infinitesimal MC steps from the scaling behavior of 
the MIPS phase transition and the 
melting transitions.

The work is organized in the following order. In \secti{sec:Model}, the 
essential elements of the kinetic MC algorithm are described, the interplay of 
persistence with interactions is illustrated on a simple case of two particles 
in one dimension (1D), and possible 
anisotropy effects in two dimensions (2D) are analyzed. 
In \secti{sec:Results}, we discuss the effect of the 
stiffness of the interparticle potential on the full quantitative phase diagram 
of two-dimensional particle systems on the activity--density plane.
The continuous-time limit of the kinetic MC dynamics is 
discussed in \secti{sec:RelevantParameters}, with a focus on the number
of relevant parameters.

\begin{table}
\caption{\label{tab:PhaseCharacteristics} Decay of correlation functions in the 
liquid, hexatic, and solid phases  in two-dimensional particle systems.}
\begin{ruledtabular}
\begin{tabular}{llcr}
Order & Liquid phase & Hexatic phase& Solid phase\\
\hline
\begin{tabular}[c]{@{}l@{}}Positional\end{tabular}    & 
short-range\footnote{$\propto \exp{(-r/\xi)}$\label{foot:exp}} & 
short-range\textsuperscript{\ref{foot:exp}}      & 
quasi-long-range\footnote{$\propto r^{-\alpha}$\label{foot:alg}} \\
\begin{tabular}[c]{@{}l@{}}Orientational\end{tabular} & 
short-range\textsuperscript{\ref{foot:exp}} & 
quasi-long-range\textsuperscript{\ref{foot:alg}}  & 
long-range\footnote{$\propto$ constant}      
\end{tabular}
\end{ruledtabular}
\end{table}

\section{Model: kinetic MC\label{sec:Model}}
\begin{figure}
\centering
\includegraphics[width=.45\textwidth]{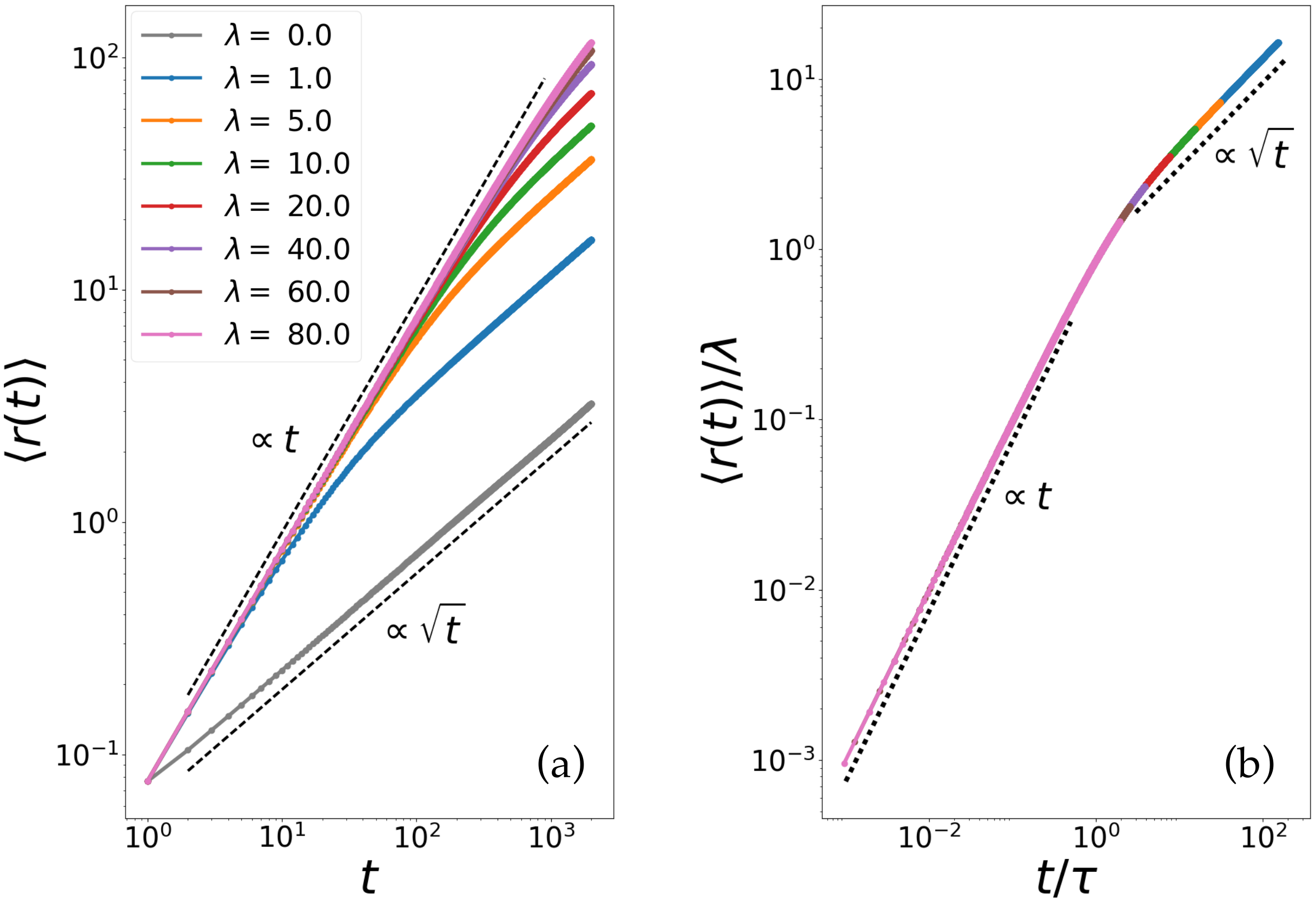}
\caption{Mean absolute displacement \emph{vs.} time $t$ for a 
single particle. \subcap{a} The crossover 
from ballistic to diffusive motion shifts to larger $t$ with increasing 
activity, measured in terms of the persistence length $\lambda$ (see 
\eq{EQ:lambda}). \subcap{b} Data 
collapse illustrating the crossover from $\propto t$ to 
$\propto \sqrt{t}$ at around $(1,1)$ expressed using 
\eqtwo{EQ:lambda}{EQ:tau}.}
\label{F:MeanAbsDisp}
\end{figure}
The kinetic MC algorithm with which we model active  dynamics 
consists of the standard Metropolis filter combined
with a memory term for the proposed moves.
The memory term is characterized by a time scale $\tau$, which  allows for a 
smooth
tuning from a passive motion (described by equilibrium statistical 
mechanics) to 
a self-propelled/persistent particle motion, where a single particle
moves ballistically (mean square displacement $\propto t^2$) for times $t \ll
\tau$ and moves diffusively (mean square
displacement $\propto t$) for times $t \gg \tau$ 
(see \subfig{F:MeanAbsDisp}{a}).

In contrast to active Brownian particles
\cite{ReviewActiveBrownian}, the velocity amplitude fluctuates in the MC 
dynamics. The dynamics
is comparable with the active Ornstein--Uhlenbeck process (see \textit{e.g.}
Ref.~\onlinecite{HowFarOrnsteinUhlenbeck}), where the velocity performs a
random walk in a harmonic potential. Similarly, in the discrete-time kinetic 
MC approach the increment performs a random walk in a box with 
reflecting boundary conditions. For a single particle in two dimensions, the 
$x$- 
and $y$-components of the increment $(\epsilon_x(t), \epsilon_y(t))$ at time 
$t$ 
are sampled from two independent Gaussian distributions of standard deviation 
$\sigma$ and the mean corresponds to the previously sampled increment 
($\epsilon_x(t-1)$ or $\epsilon_y(t-1)$, respectively). This walk is confined 
in $x$ and $y$ by reflecting boundaries at $x = \pm \delta$ and $y = 
\pm 
\delta$. In two dimensions the average distance 
a single particle covers before changing the direction is given 
by\cite{KlKaKr2018} the 
persistence length
\begin{equation}
\lambda \simeq 0.62 \frac{\delta^3}{\sigma^2}\,,
\label{EQ:lambda}
\end{equation}
and the persistence time is given by
\begin{equation}
\tau = \frac{8}{\pi^2}\frac{\delta^2}{ \sigma^2}\,.
\label{EQ:tau}
\end{equation}
This characteristic length and time scales are confirmed in numerical 
simulations. For example, the crossover from ballistic to diffusive motion in 
\subfig{F:MeanAbsDisp}{b} appears in the rescaled time-dependence of the mean 
absolute displacement  around the point $(t/\tau = 1,\langle 
r(t)\rangle/\lambda 
= 1)$.

In the many-particle case, the increment for each particle 
performs its proper random $\vec\epsilon$-walk, independent of the other 
particles. Interactions between particles are introduced by the 
Metropolis filter. In one kinetic MC step, a particle $i$ is 
chosen at random. The change of its position ($\vec{r}_i(t+1) = \vec{r}_i(t) + 
\vec{\epsilon}_i(t)$) is accepted with probability
\begin{equation}
P(E' \to E) = \min\left[1,e^{-\beta\Delta E)}\right]\,,
\label{EqMetropolisFilter}
\end{equation}
where $\Delta E = E - E'$ is the total energy change caused by the particle 
displacement $\vec{r}_i(t) \rightarrow \vec{r}_i(t+1)$. The parameter 
$1/\beta = k_\text{B}T$ is now an energy scale rather than a temperature.
The random 
$\vec\epsilon$-walk persists whether the resulting displacement is accepted or 
not. For a detailed description of the kinetic MC approach, see 
Ref.~\onlinecite{KlKaKr2018}. 
\subsection{1D Model for Persistence\label{sec:EffectOfPersistence}}
Kinetic MC, \emph{via} the memory present in its displacements, 
generates persistence in a manner that differs from equilibrium systems.
This has far-ranging consequences for many-particle systems, 
but the effects of a memory term in the equations of motion can already be 
studied for $N=1$ or $N=2$ particles.
Here we study the case of two 
particles on a ring (a line of length $L$ with periodic boundary conditions), 
interacting  with an inverse-power-law pair potential that we will use 
throughout 
this work
\begin{equation}
U(r) = u_0 \left( \frac{\gamma}{r} \right)^n\,,
\label{eq:Potential}
\end{equation}
where $\gamma$ reduces to the particle diameter in the hard-disk limit $n\to 
\infty$. 
The 1D inter-particle distance $r$ in \eq{eq:Potential} 
is easily generalized to higher dimensions.

\begin{figure}
\centering
\includegraphics[width=.45\textwidth]{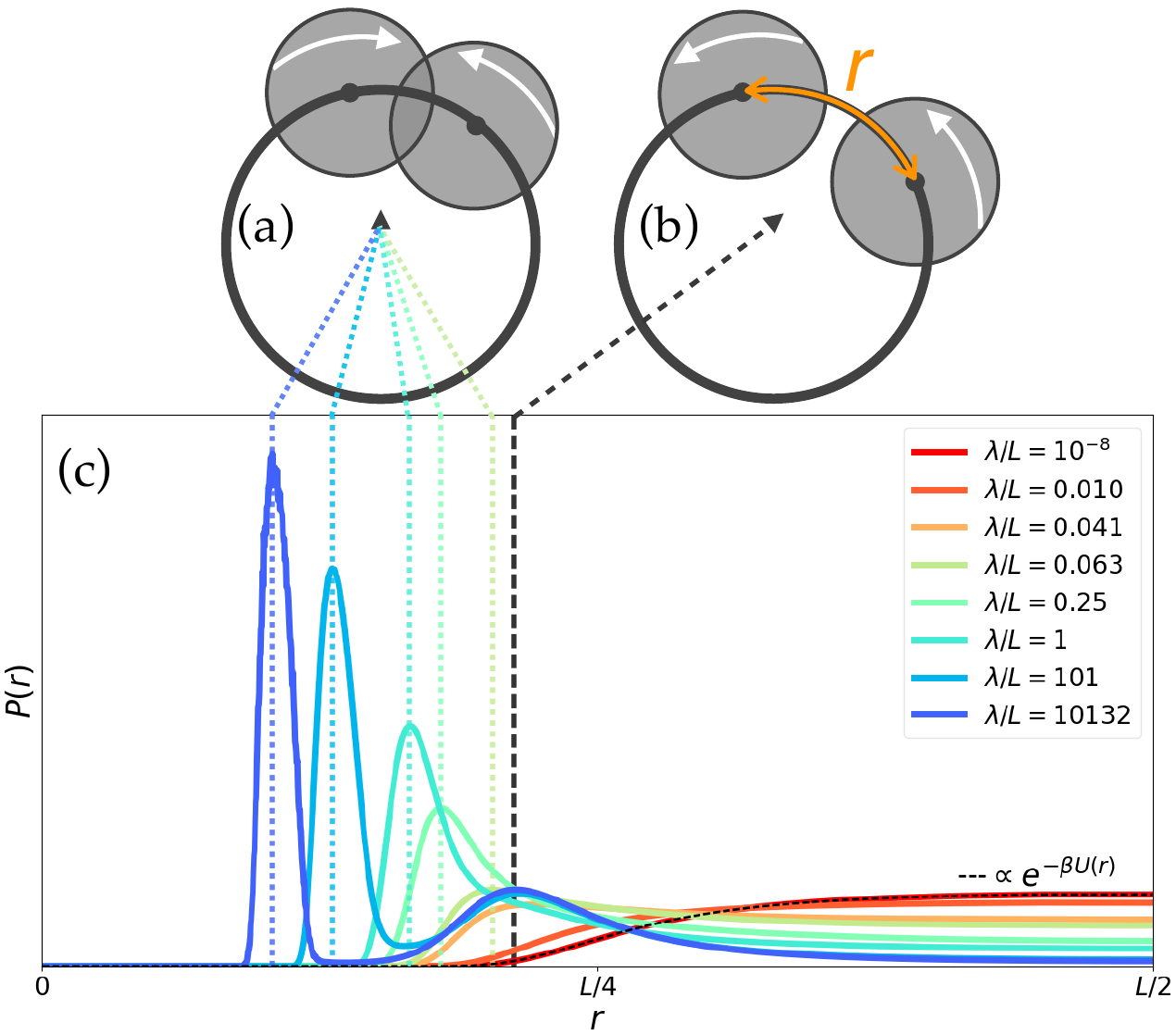}
\caption{Persistence for two 1D particles on a ring (line of length $L$ with 
periodic boundary conditions). \subcap{a} and \subcap{b} indicate the two
states of the system that lead to the  
local maxima of the pair-correlation function. White arrows indicate the sign of
the displacements $\epsilon_i$. \subcap{c} Pair 
correlation function $P(r)$ for different $\lambda$. 
Maximum jump length $\delta= L/40$
$\lambda$ is varied by changing $\sigma$. 
Data for $n = 6$, $u_0\beta = 1$. The unit length scale is set by $\gamma$ in 
\eq{eq:Potential}.}
\label{F:TwoParticlesOnRing}
\end{figure}

\fig{F:TwoParticlesOnRing} shows the probability distribution $P(r)$ of the
inter-particle distance. The maximum jump length $\delta$ is kept constant
and the persistence length $\lambda$ is varied by changing $\sigma$. As the 
interaction is repulsive, the two particles repel each other for small
$\lambda$, 
and the Boltzmann weight is maximal at $r= 
L/2$  for $\lambda = 0 $ (see
\subfig{F:TwoParticlesOnRing}{c}). 
For increased $\lambda$, a
peak appears at small $r$, and its position decreases with 
increasing
$\lambda$. This means that particles are more probable to be near each other,
which is due to the case, where the particles try to move against each
other (see \subfig{F:TwoParticlesOnRing}{a}). The higher the persistence
length, the stronger the force the particles push against the interparticle
potential barrier and therefore, the peak position shifts. This shows that
the self-propulsion force increases with $\lambda$ or equivalently with the
persistence time $\tau$, and the shift is a result of the larger number of
attempts in the Metropolis filter to increase the total energy.

At finite $\lambda$, the original (Boltzmann) peak at $r = L/2$ shifts to 
smaller $r$ and takes on a 
position that is independent of $\lambda$. This peak appears due to
arrangements where one particle \quot{hunts} the other circling around the
ring with $|\epsilon_\text{hunter}| > |\epsilon_\text{hunted}|$ (see
\subfig{F:TwoParticlesOnRing}{b}). For this to have a  non-negligible
probability, the role of the slower and faster particle has to be stable for
a sufficient time span, which explains that the peak appears only after 
overcoming a certain threshold in $\lambda$. The independence of the 
peak position on $\lambda$
is understood from the following argument. Moves of the slower particle are
always accepted by the Metropolis filter, as they decrease the total energy. In
contrast, moves of the faster particle are often rejected 
as
they increase the total energy. This leads to a competition where the slower
particle increases $r$ in every attempt, and the faster particle tries to
decreases $r$, but does not succeed in every attempt, thereby leading to
an average "hunting distance" $r$, which is independent of $\lambda$ (see
\subfig{F:TwoParticlesOnRing}{c}).

The bimodal probability distribution of $P(r)$ (see \fig{F:TwoParticlesOnRing}) 
is a consequence of the non-constant
velocity amplitude and thus in contrast to \textit{e.g.} active Brownian
particles. The result clearly shows that the self-propulsion force in the
kinetic MC dynamics depends on the persistence time $\tau$. This
is not the case in the Langevin approaches of active Brownian particles
or the active Ornstein--Uhlenbeck process. However, as discussed in
\secti{sec:RelevantParameters}, the persistent length $\lambda$ is the
relevant measure for activity is this dynamics.

\subsection{Anisotropic effects}
\begin{figure}
    \centering
    \includegraphics[width=0.475\textwidth]{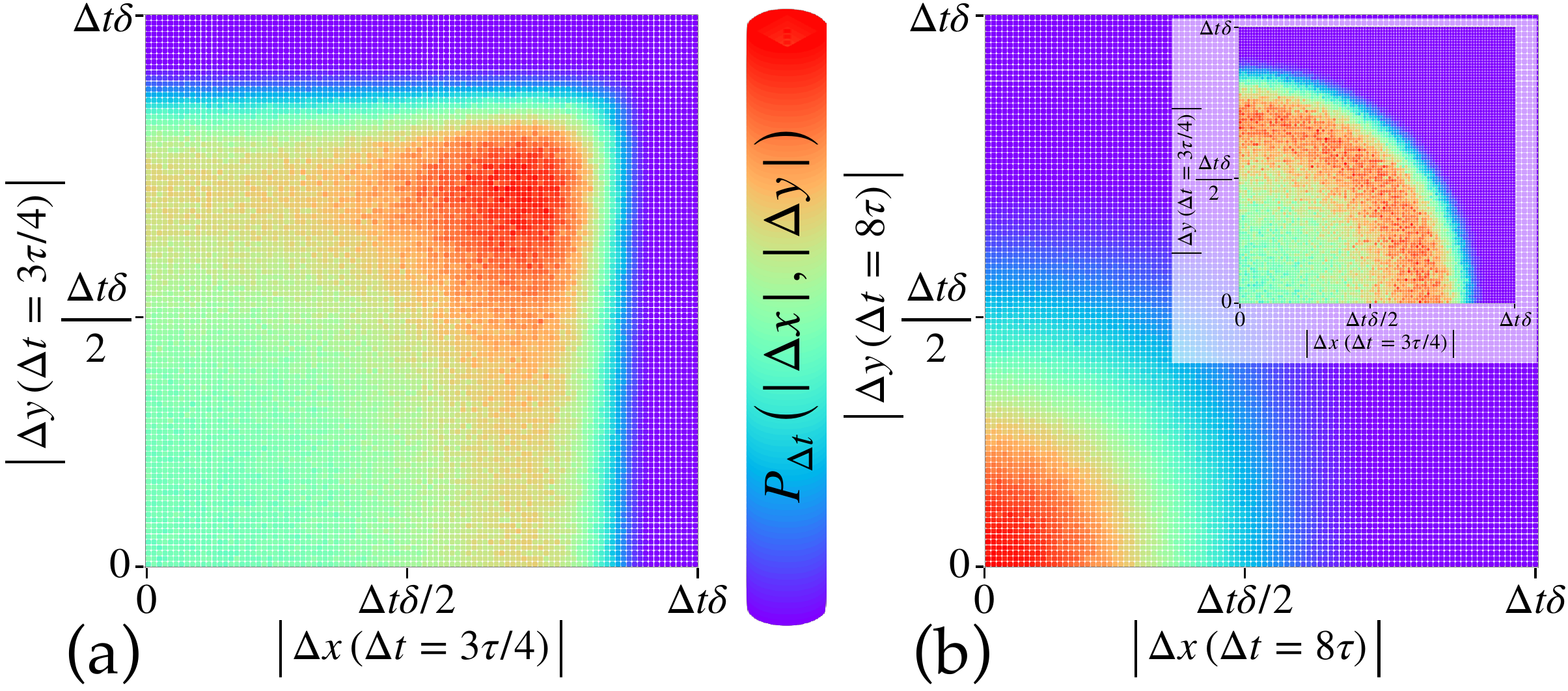}
\caption{Kinetic-MC displacement distribution $P_{\Delta t}(|\Delta x|,|\Delta 
y|)$, in time $\Delta t$, for a single particle (infinite system).
\subcap{a}: Anisotropic total displacement for small times 
($\Delta t = \tfrac{3}{4}\tau$). \subcap{b}: Isotropic total displacement 
for large times ($\Delta t = 8\tau$). Inset in \subcap{b}: 
Displacement distribution for a circular reflecting boundary for small times 
($\Delta t = \tfrac{3}{4}\tau$). (Linear color code with zero at purple.)}
\label{F2DAnisotropie}
\end{figure}
In our kinetic MC algorithm, displacements $\epsvec = (\epsilon_x, \epsilon_y) 
\in [-\delta,\delta]^2$ are confined to a square box 
rather than being sampled from an isotropic distribution (as for example a 
circle of radius $\delta$). (The 2D Gaussian distribution of the 
Ornstein--Uhlenbeck process is also isotropic.) Although the square box is 
chosen for simplicity only, it is useful to check that it does not induce 
anisotropies. This trivially follows in the passive limit for the steady-state 
probability distribution because of the detailed-balance condition. 

At small times ($t \ll \tau$), the reflecting boundary
conditions for the sampling scheme of the displacements 
introduces some degree of anisotropy in the two-dimensional single-particle
dynamics (see \subfig{F2DAnisotropie}{a}).
The
probability distribution $P_{\Delta t}(|\Delta x|,|\Delta y|)$ of the
absolute particle displacement $\Delta x$ (or $\Delta y$) in the $x$ (or $y$)
component in a time $\Delta t$ is anisotropic for $\Delta t < \tau$
(see \subfig{F2DAnisotropie}{a}) whereas, without the square box, this 
displacement (which is constructed from identically distributed, independent 
Gaussians in both dimensions) would be isotropic.
However, the isotropy is reinstalled for $t \gg \tau $ (see 
\subfig{F2DAnisotropie}{b}). As $\tau =0$ in the passive limit, the particle 
dynamics is isotropic for all times\cite{BookKrauth}.
The anisotropy  for $t < \tau$ could also be 
avoided by choosing a circular sampling box of radius $\delta$ with
reflective boundary conditions for the displacements (see
\subfig{F2DAnisotropie}{b}).  However, such a choice would be more costly to 
implement. 

\begin{figure}
	\centering
	\includegraphics[width=0.45\textwidth]{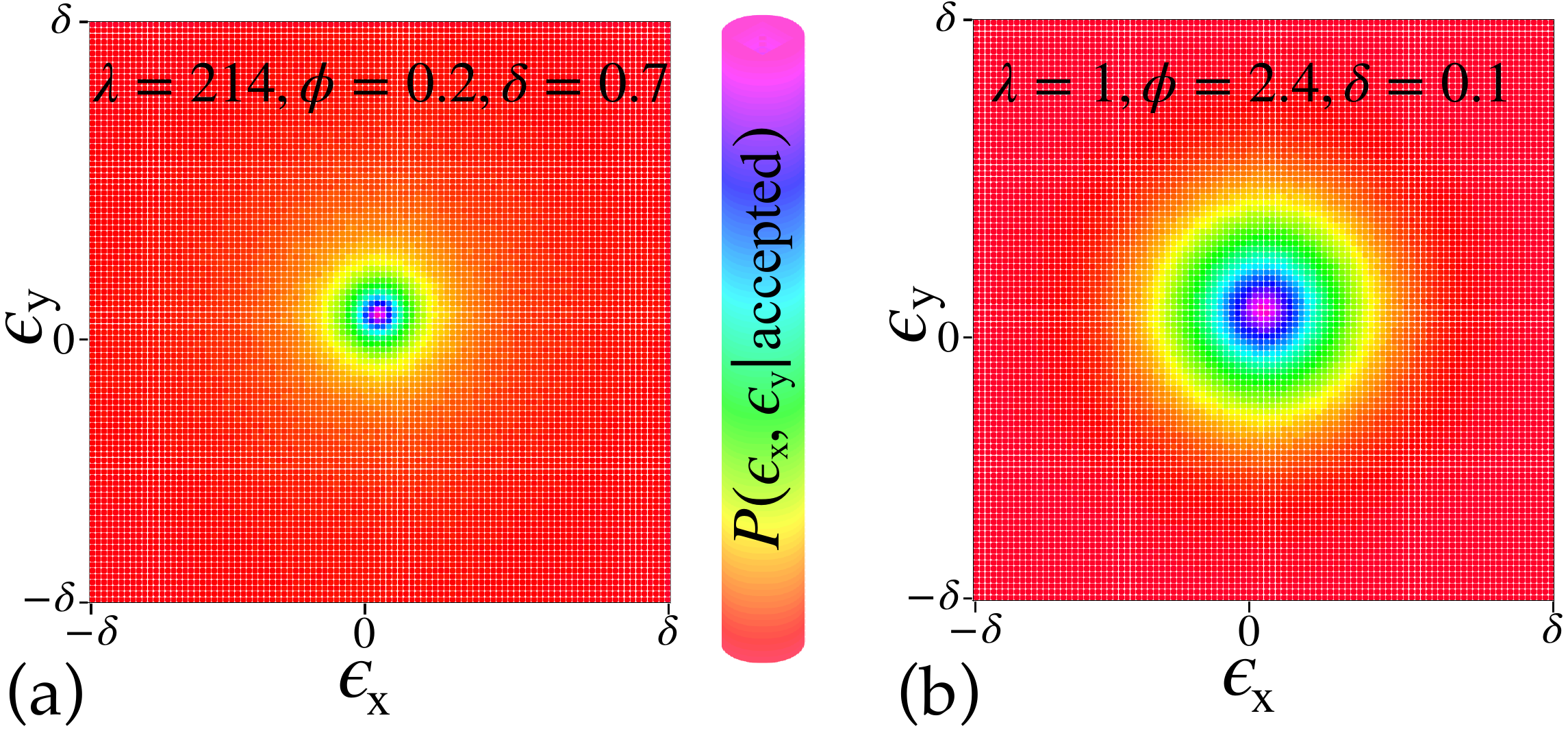}
	\caption{ Two-dimensional probability distribution of the accepted 
displacements $P(\epsilon_x, \epsilon_y | \text{accepted})$.
\subcap{a} Many-body system in the MIPS region. \subcap{b} 
In the solid near the 
solid--hexatic transition at a density far above the 
equilibrium melting lines ($n = 6$, $N= 10976$).}
       \label{F2DAcceptedDispl}
\end{figure}

\begin{figure}
\centering
\includegraphics[width=0.475\textwidth]{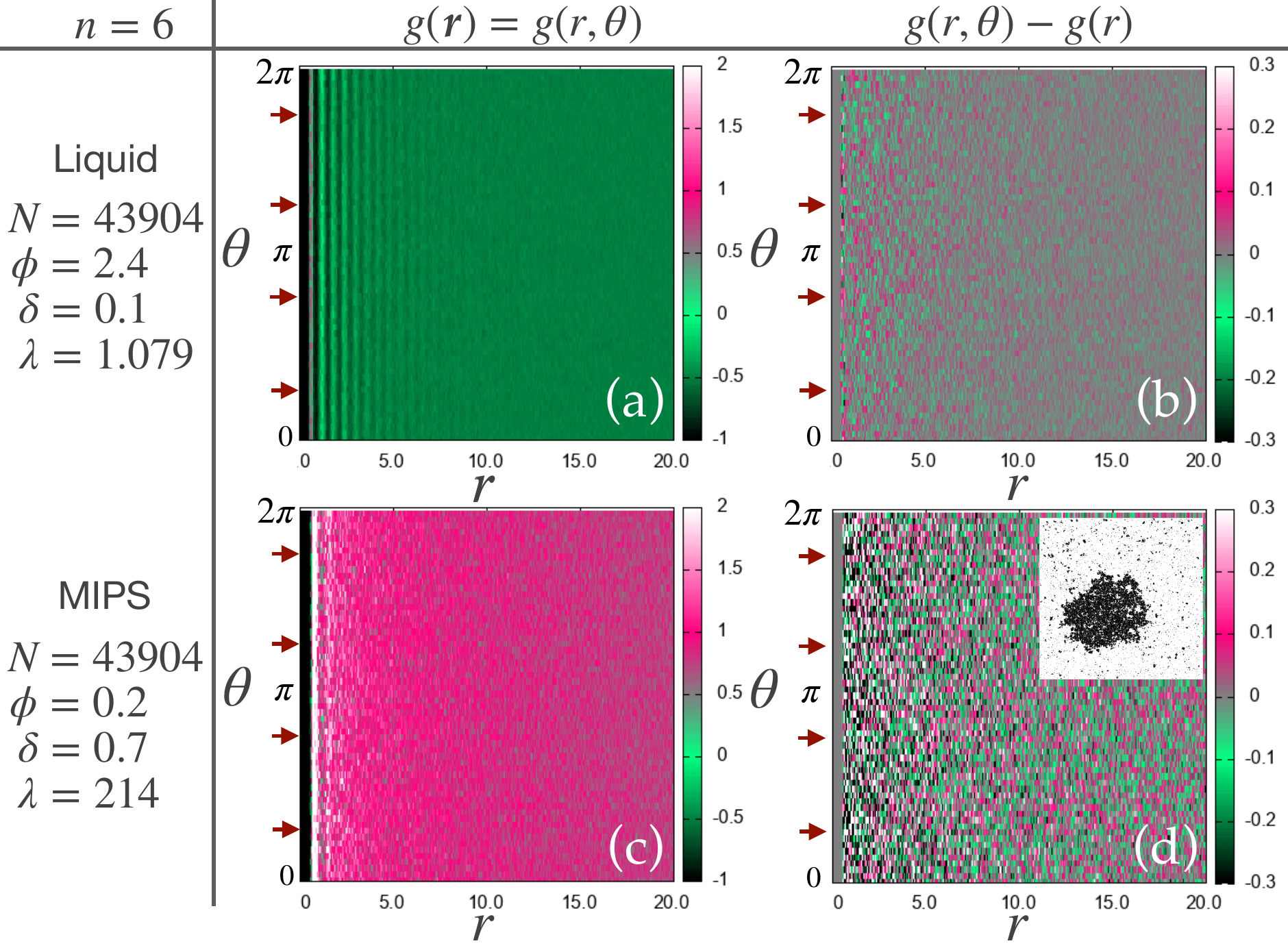}
\caption{Effective isotropic dynamics 
($n = 6$, $N= 43904$).
\subcap{a} and \subcap{c}: pair-correlation function $g(r,\theta)$
in polar coordinates averaged over $100$ configurations. \subcap{b} and 
\subcap{d}:
difference between $g(r,\theta)$ and its angular average $g(r)$. The red arrows 
indicate $\pi/4, 3\pi/4, 5\pi/4$ and
$7\pi/4$. Inset in \subcap{d}: snapshot of configuration showing MIPS
corresponding to \subcap{c} and \subcap{d}.}
\label{F: 2DPairCorrAnisoSolHexLiq}
\end{figure}

We conjecture that the square box in fact renders anisotropic
the long-time dynamics neither for $N=1$ nor for the many-body case.
In the dilute case (where $\lambda$ is much smaller than the mean free path)
the kinetic MC dynamics effectively reverts to the detailed-balance
dynamics as interactions between particles happen at the diffusive time
scale.  At higher densities, anisotropy in the many-body properties
might arise if the probability distribution of the accepted displacements 
is itself anisotropic. However, this is not the case (see 
\fig{F2DAcceptedDispl}).
At higher densities, all large proposed displacements have a vanishing 
probability to be accepted by the Metropolis filter, thus leading to an 
effectively isotropic dynamics. Therefore, the Metropolis filter effectively 
realizes a circular reflecting $\epsvec$-sampling box without additional 
computational cost. Pair-correlation functions are also found to be perfectly 
isotropic at high density, both in the 
motility-induced liquid phase (at a density deep inside the equilibrium solid
phase) and in the MIPS region (see \fig{F: 2DPairCorrAnisoSolHexLiq}).

\section{Two-dimensional simulation results\label{sec:Results}}
Our simulations are performed in an ensemble of $N$ particles confined 
to a rectangular\cite{KlKaKr2018} box of volume $V$ with periodic 
boundary conditions. The  density $\phi = \gamma^2 N/V$ (with 
$\gamma$ from \eq{eq:Potential}) is varied by changing $V$.
In the simulations $\delta$ is kept constant and the activity is varied by 
changing $\sigma$.

\begin{figure}
\centering
\includegraphics[width=0.45\textwidth]{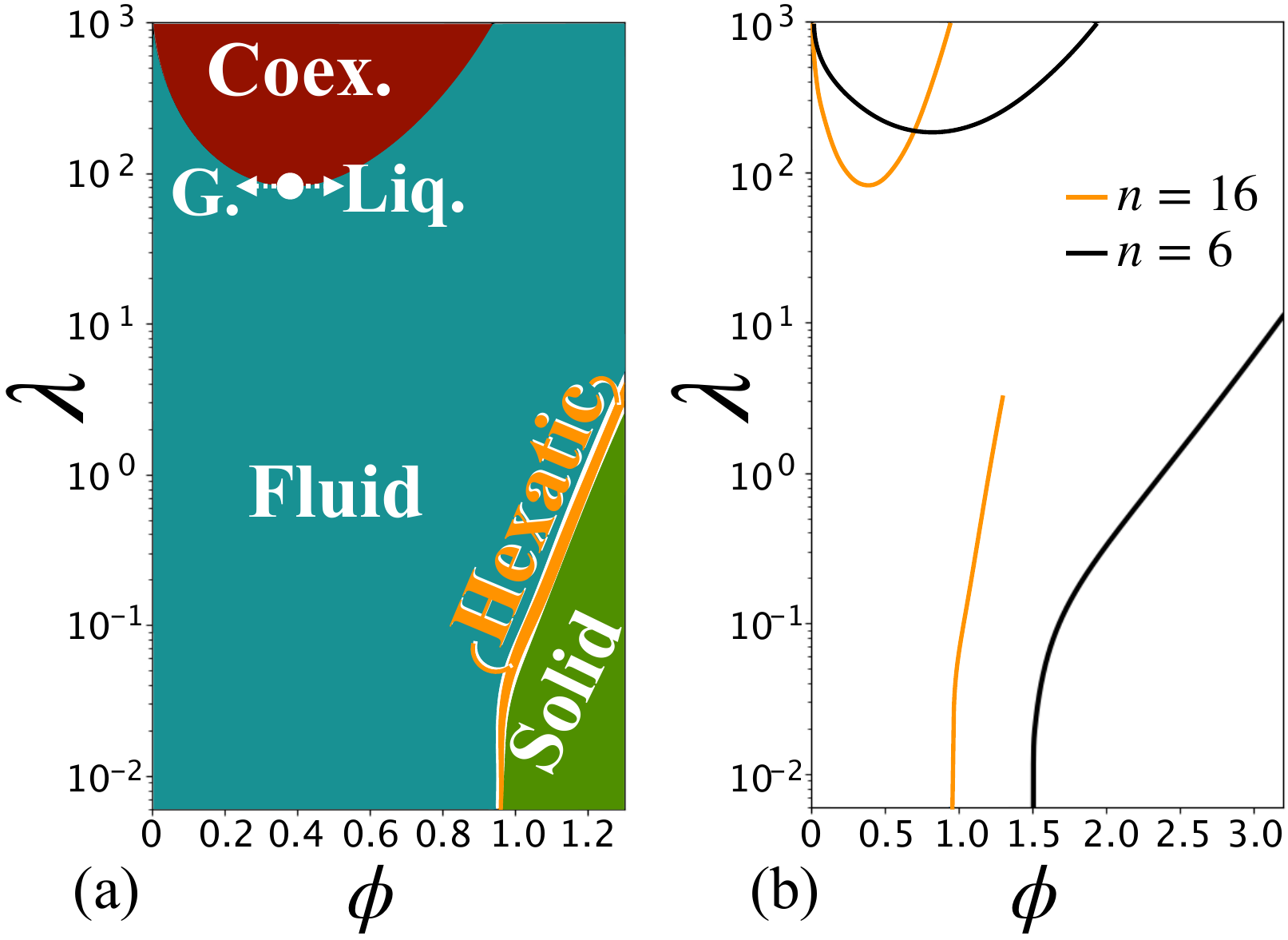}
\caption{Phase diagram as a function of density $\phi$ and persistence 
length $\lambda$
\subcap{a}: Soft-disk potential 
with $n = 16$. \subcap{b}: Comparison of phase boundaries \cite{KlKaKr2018} 
for $n=6$ with the steeper $n=16$ case
($\delta = 0.1$, $\gamma = 1$, $u_0\beta = 1$).
MIPS is always separated from the melting transitions. }
\label{F:FullPhaseDiagram}
\end{figure}

\subsection{Full phase diagram and the effect of 
stiffness\label{sec:PhaseDiagram}}
\subfig{F:FullPhaseDiagram}{a} shows the full phase diagram for the
potential in \eq{eq:Potential} with $n = 16$ on the $\phi$--$\lambda$ plane. At
all $\lambda$, the equilibrium two-step melting 
transition\cite{KapferKrauth2015} is recovered. For 
increasing $\lambda$, the
melting lines shift to higher densities.
Remarkably, the hexatic phase 
separating the liquid and solid phases is stable even far
from equilibrium. This non-equilibrium two-step melting can be induced
either by reducing the density (just as in equilibrium) or by increasing
the persistence length.  In addition to these melting transitions, 
at low $\phi$ but high $\lambda$, a motility-induced liquid--gas 
coexistence region opens up. It is separated form the melting transitions by a 
disordered fluid phase. This generalizes the phase diagram under the same 
dynamics, but for $n = 6$, found previously\cite{KlKaKr2018}. A change of $n$ 
(see \eq{eq:Potential}) only shifts the positions of the phase boundaries
(see \subfig{F:FullPhaseDiagram}{b}).
As for the equilibrium melting transitions
\cite{KapferKrauth2015}, both the liquid--hexatic and the hexatic--solid phase 
boundaries shift at constant persistence length $\lambda$ to smaller
densities with increasing $n$.

Increasing $\lambda$ shifts the melting transitions to higher
densities. However, the shift is smaller for larger $n$, resulting in
steeper transition lines (see \subfig{F:FullPhaseDiagram}{b}).  At the
same time, the onset of the motility-induced liquid--gas coexistence shifts
to smaller $\phi$ and $\lambda$ and the coexisting region shrinks. This
ensures that the liquid--gas coexistence and the melting transitions
remain disjoint. In \secti{sec:MIPS}, we argue that there is no
singular change in the phase diagram even in the hard-disk limit $n\to
\infty$.  

\subsection{Non-equilibrium two-step melting \label{sec:Melting}}
In equilibrium, the Mermin--Wagner theorem forbids long-range translational
(\emph{i.e.} crystalline) order in a 2D particle system with 
short-range 
interactions \cite{MerminWagner,Mermin1968}. 
However, at large densities,
particles can arrange in locally hexagonal configurations. This can lead to 
two different high-density
phases\cite{NelsonHalperin1979,HalperinNelson1978,Young1979}, 
which are characterized 
by different degrees of orientational and positional order (see 
\tab{tab:PhaseCharacteristics}). In this section, we 
define these measures of order and use them to quantify the two-step
melting far from equilibrium.

The local bond-orientational order parameter $\psi_6(\bm{r}_i)$ measures the 
six-fold orientation near a particle $i$. It is defined as
\begin{equation*}
\psi_6(\bm{r}_i) = \frac{1}{\text{number of neighbors $j$  of  $i$ } }
\sum_j 
\exp(6 \imath\theta_{ij})\,,
\end{equation*}
where $\imath$ is the imaginary unit and $\theta_{ij}$ it the angle enclosed by 
the $x$-axis and the connection line between particle $i$ and its neighbor $j$. 
Here, we use the Voronoi construction to identify neighbors and 
$\psi_6(\rvec_i)$ is calculated with Voronoi 
weights\cite{VoronoiWeights}. Then, the correlation function
\begin{equation}
g_6(r)\propto \left\langle \sum_{i,j}^N \psi^\star_6(\bm{r}_i)\psi_6(\bm{r}_j) 
\delta(r - r_{ij}) \right\rangle
\end{equation}
is a measure of the correlation of the local six-fold orientational order at 
distance $r$ and its decay is used to quantify the degree of orientational 
order 
in the system (see \tab{tab:PhaseCharacteristics}).

The direction-dependent pair-correlation function $g(x,y)$ provides
a measure for the positional order. This two-dimensional
histogram is averaged over different configurations $C$ after re-aligning
\cite{BernardKrauth2011} $g_C(x,y)$ such that the $\Delta$x-axis points
in the direction of the global orientation parameter $\Psi_6(C) = \sum_i^N
\psi_6(\bm{r}_i)$ of $C$. Then, the decay of, \textit{e.g.} $g(x,0)$ determines
the degree of positional order.

The correlation functions ($g(x,0)$ and $g_6(r)$), allow one
to identify the two-dimensional phases by the properties summarized
in \tab{tab:PhaseCharacteristics}. In equilibrium, the 
Kosterlitz--Thouless--Halperin--Nelson--Young theory
provides an additional selection 
criterion\cite{NelsonHalperin1979,HalperinNelson1978,Young1979}, 
where the exponent (defined in \tab{tab:PhaseCharacteristics}) $\alpha \leq 1/4$ 
for the orientational order and $\alpha \leq 1/3$ for the positional order give 
theoretical bounds for the hexatic phase. However, these bounds are not shown to 
apply outside equilibrium. We thus identify the phases by the characteristic 
decay of $g(x,0)$ and $g_6(r)$.

\begin{figure}
\includegraphics[width = 0.48 \textwidth]{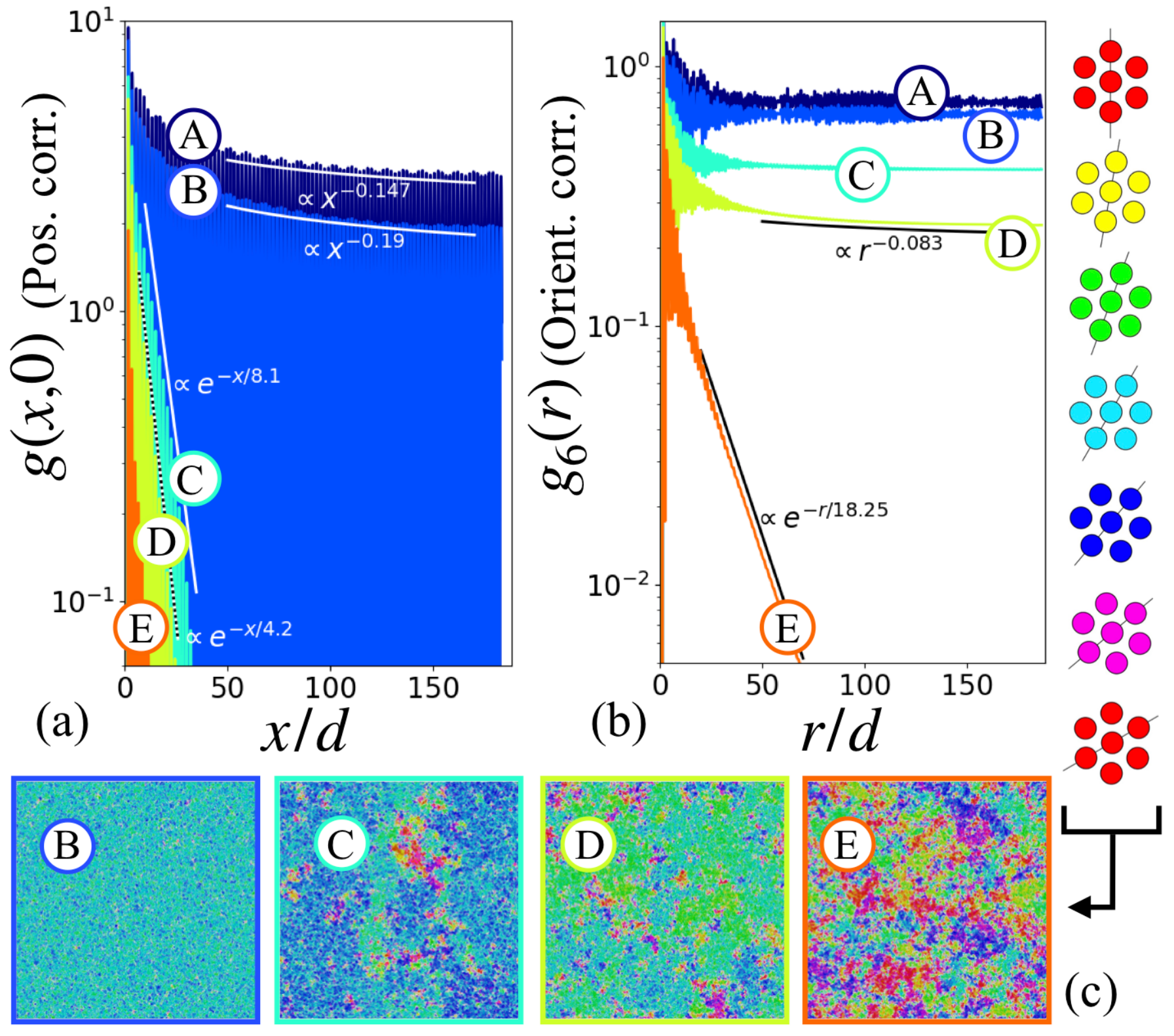}%
\caption{\label{F:Melting_n16} Two-step melting for $n=16$ at 
high density.
\subcap{a}: 
Positional correlation $g(x,y=0)$ 
\subcap{b} orientational 
correlation $g_6(r)$. 
 Particles in 
\subcap{c} are color-coded according to local orientational order 
$\psi_6$. \quot{A} and \quot{B} are solid (algebraic 
$g(x,0)$, long-range  $g_6(r)$). \quot{C} and \quot{D} are hexatic (exponential $g(x,0)$, 
algebraic $g_6(r)$). \quot{E} is liquid 
(both decays exponential).
($d = (\pi N /V)^{-1/2}$,
$\phi = 1.2$, $N 
= 43904$, $\delta = 0.1$, $u_0\beta = 1$, \quot{A}: $\lambda = 0.5$, \quot{B}: 
$\lambda = 
0.7$, \quot{C}: $\lambda = 1.0$, \quot{D}: $\lambda = 1.1$, \quot{E}: $\lambda 
= 1.4$)}
\end{figure}

Orientational and positional correlation functions change as the
system melts from solid to liquid passing through the hexatic phase
(see \fig{F:Melting_n16}, 
at a density far above the equilibrium
melting point). 
Our simulations clearly identify solid state points with 
power-law decay in $g(x,0)$ and constant $g_6(r)$, hexatic
state points with exponential decay of  $g(x,0)$ and quasi-long-range order in 
$g_6(r)$, and liquid state points, where both correlations decay 
exponentially. Snapshots illustrate these different phases (see 
\fig{F:Melting_n16} and \tab{tab:PhaseCharacteristics}).
The power-law exponent in $g_6(r)$
grows when approaching the liquid--hexatic transition and therefore,
weakening the order in the hexatic phase. This behavior of $g_6(r)$ is in
agreement with equilibrium studies\cite{KapferKrauth2015} and with our earlier
results\cite{KlKaKr2018} obtained with the same dynamics for $n = 6$.

We check that our simulations indeed reach the steady state (as in 
earlier 
work\cite{KlKaKr2018}) by verifying the convergence of the spatial correlation 
functions to the same steady state starting from a crystalline and a liquid 
initial particle arrangement.
These very time-consuming computations assure that the defining 
decays of the correlation 
functions 
obtained in the hexatic phase reflect the physical system and not a bias 
introduced by the initial condition. 

\subsection{Motility-induced phase separation\label{sec:MIPS}}

At sufficiently low densities and high activities, a liquid--gas
coexistence region opens up with a roughly U-shaped phase boundary (see
\subfig{F:FullPhaseDiagram}{a}). Following an 
analysis of local densities, 
which was applied earlier\cite{KlKaKr2018} for $n = 6$, we confirm that 
also for 
$n = 16$ the densities ($\phi_\text{Liquid}$ and $\phi_\text{Gas}$) of the two 
coexisting phases depend on $\lambda$ but not 
on the global density $\phi$. Therefore, the low-density boundary of the MIPS 
region in \subfig{F:FullPhaseDiagram}{a} is given by\cite{KlKaKr2018}
$\phi=\phi_\text{Gas}(\lambda)$ and the high-density boundary by $\phi =
\phi_\text{Liquid}(\lambda)$, respectively.

\begin{figure}
\includegraphics[width = 0.475 \textwidth]{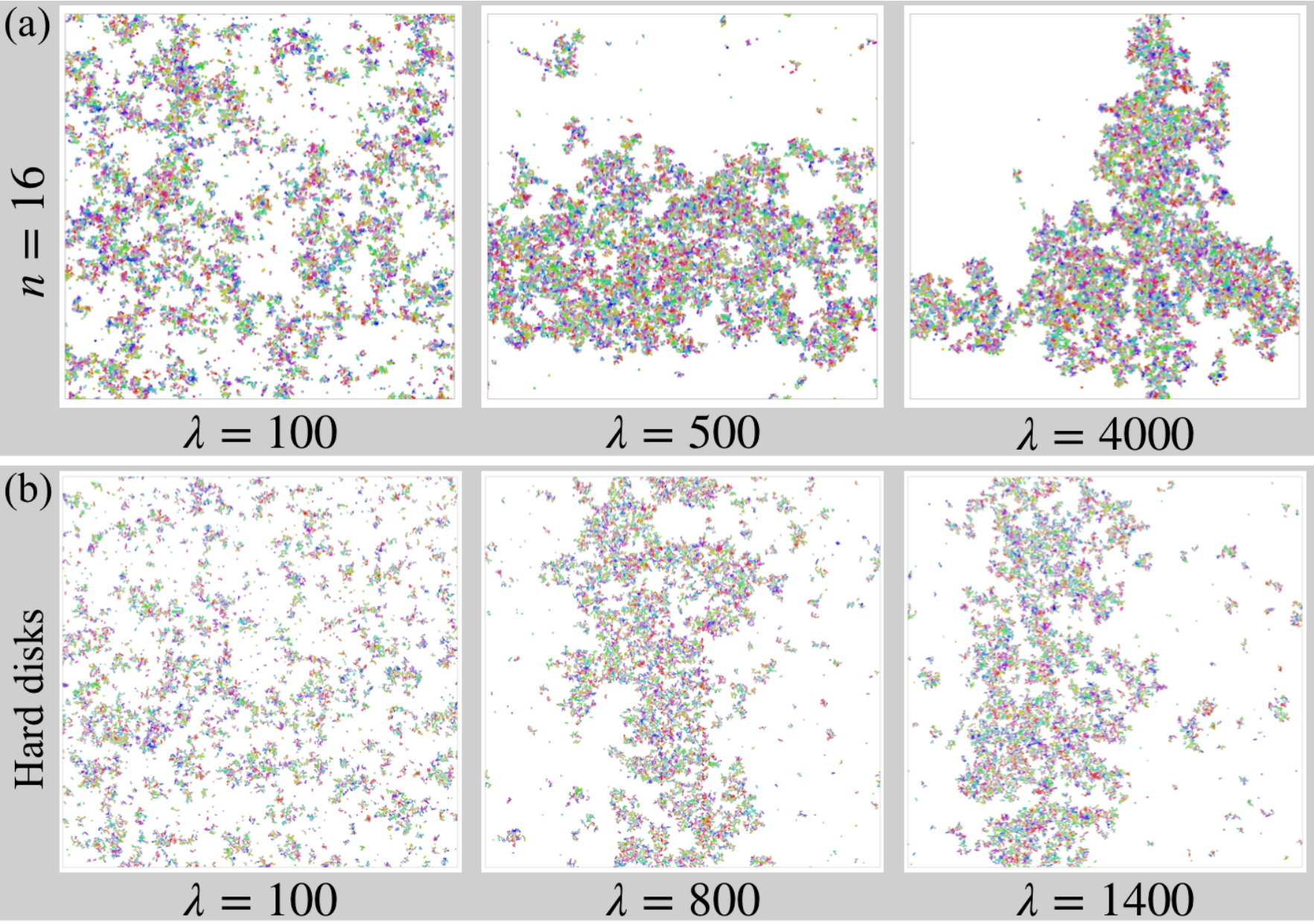}%
 \caption{\label{F:MIPS-HardSoft} Short-range order in MIPS for $n = 16$ and 
 for
hard disks. \subcap{a} Snapshots for  $n = 16$ at $\phi = 0.4$.
\subcap{b}: Snapshots for hard disks ($n \to \infty$)  at $\phi = 0.2$. The 
liquid phase is clearly identified by a short-ranged orientational correlation 
illustrated by the $\psi_6$ color code defined in \fig{F:Melting_n16}. Data for 
$N = 10976$, $\delta = 0.1$, $\beta u_0 = 1$.}
\end{figure}

A much discussed 
question\cite{FilyMarchetti2012,KineticModel,JanusParticleSpeck2013,Dumbbell1,
MeltingABP} 
concerns the nature of the two
phases at coexistence. 
We can 
clearly identify the high-density phase as liquid, and 
MIPS as a liquid--gas coexistence. The particles in the
snapshots in \fig{F:MIPS-HardSoft} are $\psi_6$-color-coded (see
\fig{F:Melting_n16} for definition), illustrating  short-ranged
orientational order in the liquid phase. We do not observe that the 
orientational correlation in the liquid phase increases 
with increasing $\lambda$. Even at very high activities (e.g. at $\lambda = 4 
\times 10^3$ in \subfig{F:MIPS-HardSoft}{a}), the local orientational order 
changes upon length scales of the order of the interparticle distance (also 
compare with \subfig{F:Melting_n16}{c} point E).

We do not observe any quantitative difference with the orientational order in 
MIPS previously observed\cite{KlKaKr2018} for $n=6$ for the same kinetic MC 
dynamics. Furthermore, we also 
recover MIPS in the hard-disk system in the form of a liquid--gas coexistence 
(see \subfig{F:MIPS-HardSoft}{b}).  We conjecture from these findings 
that the separation of the MIPS region from the melting transitions is a 
generic 
feature of self-propelled particles, at least within kinetic MC dynamics.

\section{Relevant parameters\label{sec:RelevantParameters}}
\begin{figure}
\includegraphics[width = 0.45 
\textwidth]{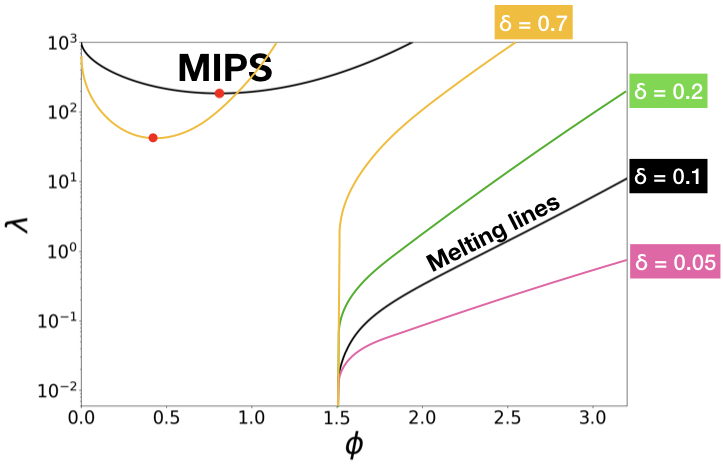}
\caption{\label{F:PhaseDiagramDeltaDependence} Phase boundaries (MIPS and 
melting transitions) for different step sizes $\delta$. Data for $n = 6$, 
$u_0\beta = 1.0$.}
\end{figure}

So far (as our previous work \cite{KlKaKr2018}), we have
considered the phase diagram as a function of the persistence length  and the 
density, keeping
the maximum step size $\delta$ constant 
(see \fig{F:FullPhaseDiagram}). However, $\delta$ 
has a profound influence on the phase boundaries 
(see \fig{F:PhaseDiagramDeltaDependence}). Keeping $\phi$ constant, the melting lines shift to 
smaller activities as $\delta$ is decreased, while in contrast 
the MIPS phase boundary shifts to larger activities. In this section, we 
study this 
$\delta$-dependence of melting and of MIPS. 

The kinetic MC dynamics depends on three parameters ($\delta$, $\lambda$, 
$\phi$). We now show that MIPS (seen at high $\lambda$) and the melting 
close to equilibrium are described by a different reduced set of relevant 
parameters in the $\delta \to 0$ limit. 
The single relevant parameter\cite{KapferKrauth2015}, which describes the 
melting transitions for  inverse-power-law potentials is not commensurate with 
the reduced parameters for MIPS, as it does not 
capture the critical melting density in the passive 
limit. Therefore, there are separate descriptions of MIPS and the melting  
transitions. 

Here we introduce the Master equation as a  stochastic descriptions of our 
kinetic MC dynamics, in addition to a Langevin description and the associated 
Fokker--Planck equation. We also compare the dimensional reduction of the 
relevant parameters with other stochastic models of active matter.

\subsection{A simple argument\label{SecApproximateDynamics}}
We first address the question of relevant parameters with a simple 
argument for a single particle. (A more detailed analysis is presented in the 
following 
 \secti{S: exact analysis of MC} and \secti{Multi-particle conti limit KMC}.) 
We consider the small-$\delta$ limit, which is justified for the choice of
parameters used in the simulations.  For a single particle in a 1D
confining potential $U(x)$, the kinetic MC rule is approximated by the 
following discrete-time ($\Dt = 0,1,2,\dots$) dynamics:
\begin{subequations}
\begin{eqnarray}
\epsilon_{\Dt+1} &=\epsilon_\Dt+r_\Dt+R\left(\frac{\epsilon_\Dt}{\delta}\right)\,,\\
 x_{\Dt+1} &= x_\Dt+\epsilon_\Dt\,f(x_\Dt,\epsilon_\Dt)\,,
\label{EQ_ApproxDynam}
\end{eqnarray}
\label{EQ_ApproxDynam_set}
\end{subequations}
where $r_\Dt$ is a Gaussian random number with $\langle r_\Dt\rangle=0$, $\langle
r_\Dt r_{\Dt'}\rangle=\sigma^2\delta_{\Dt,\Dt'}$ and
\begin{equation}
f(x,\epsilon)=\min\left\{1,\exp\left(-\frac{U(x+\epsilon)-U(x)}{k_\text{B}T}
\right)\right\}
\label{Eq:Meto1}
\end{equation}
is the acceptance rate of the Metropolis filter in 
\eq{EqMetropolisFilter}.
The reflecting boundary at $\epsilon=\pm \delta$ is denoted by $R$ (without 
specifying it rigorously) and $\delta_{\Dt,\Dt'}$ is the Kronecker delta.

Defining a set of rescaled coordinates, 
\begin{equation}
t=\Dt\, \delta, \qquad v(t)=\frac{\epsilon_\Dt}{\delta}, \qquad 
\xi(t)=\frac{r_\Dt}{\delta^2}, \qquad x(t)=x_\Dt\,,
\label{Eq:BallisticScaling}
\end{equation}
and taking the small-$\delta$ limit, we get
\begin{align}
 \dot{v}(t)&=\xi(t)+R(v(t))+\mathcal{O}(\delta)\\
\dot{x}(t)&= v(t)\,f\big(x(t),v(t)\delta\big)+\mathcal{O}(\delta)
\end{align}
where $\langle \xi(t)\rangle=0$,  $\langle \xi(t)\xi(t') 
\rangle=\lambda^{-1}\delta(t-t')+\mathcal{O}(\delta/\lambda)$,
with\footnote{The persistence length used for the numerical results 
\eq{EQ:lambda} differs by a numerical constant in both one and two 
dimensions.} $\lambda=\delta^3/\sigma^2$ and we use a 
continuous limit of the Kronecker delta 
$\delta_{\Dt,\Dt'} \simeq \delta\,\delta
(t-t')$, with $\delta(x)$ being the Dirac delta function.
Moreover, in the small-$\delta$ limit, using Taylor expansion, we get
\begin{equation*}
\frac{U(x+v\delta)-U(x)}{k_\text{B}T} = \Gamma_1 v U'(x)\left[ 1 + 
\frac{\delta v}{2}\frac{ U''(x)}{U'(x)}+\cdots\right]\,,
\end{equation*}
where $\Gamma_1=\delta/k_\text{B}T$. As long as $\delta U''(x)/U'(x) \ll 1$,
from \eq{Eq:Meto1}, we get $f[x(t),v(t)\delta]\simeq h[x(t),v(t)]$, with
\begin{equation}
h[x(t),v(t)] = \min\left\{1,\exp\left(-\Gamma_1 v U'(x)\right) \right\}\,. 
\label{eq:h def 0}
\end{equation}
This gives a continuous dynamics for $\delta \to 0$
\begin{subequations}
\begin{align}
\dot{v}(t)&=\xi(t)+ R(v(t))+\mathcal{O}(\delta)\,, \label{eq:Langevin cont v} \\
\dot{x}(t)&= 
v(t)\,h\big(x(t),v(t)\big)+\mathcal{O}(\delta)+\mathcal{O}\left(\delta\frac{
U''(x)}{U'(x)}\right)\,.
\label{eq:Langevin cont x}
\end{align}
\label{eq:contiLangevin1}
\end{subequations}
Clearly, this rescaled dynamics depends on only two relevant 
parameters, namely
\begin{equation}
\Gamma_1 = \frac{\delta}{k_\text{B}T}\,, \text{ and  } \lambda = 
\frac{\delta^3}{\sigma^2}\,. 
\label{Eq:reducedParameters1}
\end{equation}
However, for this description to be valid, the subleading terms have to be
negligible, leading to the following range of validity for 
the scaling in \eq{Eq:reducedParameters1}:
\begin{itemize}
\item[a)]
 $\delta\ll \lambda$ (from $\langle\xi(t)\xi(t')\rangle$),
\item [b)]
$\delta\ll \lambda^{-1/2}$ (from \eq{eq:Langevin cont v}), and
\item[c)]
 $\delta\ll 1$ and $\delta \ll U'(x)/U''(x)$ (from \eq{eq:Langevin cont 
x}).
\end{itemize}
Therefore, the scaling expressed in the two-parameter
reduction in \eq{Eq:reducedParameters1} breaks down in a) the passive limit,
b) at very high persistence lengths (for constant $\phi$), 
and c) at high density (for constant $\lambda$).

\subsection{Stochastic description of the MC dynamics\label{S: exact analysis 
of MC}}
We begin the rigorous analysis by considering a single 
particle in a 1D confining potential $U(x)$. The kinetic MC 
dynamics is Markovian in the $(x,\epsilon)$ space. The 
discrete kinetic MC time is denoted by $\Dt=0,1,2,\dots$. The 
conditional probability for a transition $(y,\epsilon')\rightarrow 
(x,\epsilon)$ 
in one time step is given by the Markov matrix
\begin{equation}
M(x,\epsilon | y,\epsilon') = g(\epsilon,\epsilon')W_{\epsilon}(x,y)\,,
\label{eq:Master cond prob}
\end{equation}
where, $\epsilon$ is sampled with probability $g(\epsilon,\epsilon')$
and $W_{\epsilon}(x,y)$ is due to the Metropolis filter. It can be 
shown\cite{KlamserThesis} that
\begin{align}
g(\epsilon,\epsilon') 
=\frac{1}{2\delta}+&\frac{1}{\delta}\sum_{m=1}^{\infty}\exp\left[-\frac{\pi^2 
\sigma^2 m^2}{8\delta^2}\right]\cr
\cos&\left(\frac{m 
\pi}{2\delta}(\epsilon+\delta)\right)\cos\left(\frac{m 
\pi}{2\delta}(\epsilon'+\delta)\right),
\label{DisplacementDistribution}
\end{align}
whereas the Metropolis filter in \eq{EqMetropolisFilter} yields
\begin{equation}
W_{\epsilon}(x, 
y)=f(y,\epsilon)\,\delta(x-y-\epsilon)+[1-f(y,\epsilon)]\,\delta(x-y), 
\label{CondiPropPos}
\end{equation}
with $f(x,\epsilon)$ as defined in \eq{Eq:Meto1}. Using this in the 
corresponding discrete-time Master equation
\begin{equation*}
P_{\Dt+1}(x,\epsilon) = \int dy \int_{-\delta}^\delta d\epsilon' \, M(x,\epsilon
| y, \epsilon') P_\Dt(y,\epsilon')
\end{equation*}
gives 
\begin{align}
P_{\Dt+1}(x,\epsilon) =\int d\epsilon' g(\epsilon , \epsilon')P_\Dt(x,\epsilon')
+\int d\epsilon' g(\epsilon ,
\epsilon')\cr
\left\{f(x-\epsilon,\epsilon)P_\Dt(x-\epsilon,\epsilon')-f(x,\epsilon) 
P_\Dt(x,\epsilon')\right\}\,. \label{eq:Master discrete}
\end{align}
This describes the exact time evolution of the probability $P_\Dt(x,\epsilon)$
in the kinetic MC dynamics\footnote{The approximate dynamics in
\eq{EQ_ApproxDynam_set} has a different Master equation.
Nevertheless, both describe the
same dynamics in the small-$\delta$ limit.}.
In the passive limit, this satisfies the standard detailed-balance condition 
with respect to the Boltzmann distribution (see 
\app{S:PassiveDiffusive}). 

\subsubsection*{Rescaled coordinates}
To determine the relevant number of control parameters, we
use the scaled coordinates defined in \eq{Eq:BallisticScaling} in the Master 
equation (\eq{eq:Master discrete}):
\begin{align}
\tilde{P}_{t+\delta}(x,v) = &\int dv' \tilde{g}(v , v')\tilde{P}_t(x,v')\cr
& +\int dv' \tilde{g}(v , v')\left\{f(x-\delta v,\delta v)\tilde{P}_t(x-\delta 
v,\delta v')\right.\cr
&\left. -f(x,\delta v) \tilde{P}_t(x,\delta v')\right\}\,,
\label{EQ_rescaled_Master_1}
\end{align}
with $P_\Dt(x,\epsilon) = \tilde{P}_t(x,v) / \delta$ and 
$g(\epsilon,\epsilon') = \tilde{g}(v,v') / \delta$.

We have shown in \secti{SecApproximateDynamics} that the effective number
of control parameters can be reduced by taking the small-$\delta$ limit. In 
this 
limit $f(x,\delta v) \simeq h(x,v)$ (see \eq{eq:h def 0}). Using 
a Taylor expansion, this leads to
\begin{align*}
\frac{\partial}{\partial t}\tilde{P}_{t}(x,v) = &a_1(v)\partial_v 
\tilde{P}_t(x,v)+\frac{a_2(v)}{2}\partial_v^2 \tilde{P}_t(x,v) \cr
&-\int dv' \tilde{g}(v , v')v \frac{\partial}{\partial x}\left\{h(x, 
v)\tilde{P}_t(x, 
v')\right\}+\cdots\,,
\end{align*}
where we used $\int dv'\, 
\tilde{g}(v,v')=1$ and
\begin{equation*}
a_m(v)=\frac{1}{\delta}\int dv' (v'-v)^m \tilde{g}(v,v')\,.
\end{equation*}
$a_m(v)$ can be computed using 
\eq{DisplacementDistribution}. Alternatively, we can use  
the free case
\begin{equation}
\tilde{g}(v,v') =\frac{1}{\sqrt{2\pi 
\delta / \lambda
} 
}
\exp\left[-\frac{\left(v-v' 
\right)^2}{2 \delta / \lambda}\right]
\label{DisplacementDistribution free}
\end{equation}
in combination with a zero-current condition on $\tilde{P}_t(x,v)$ for
the reflecting boundary. This simplifies the calculation of $a_m(v)$,
giving  $a_1(v)=0$, and
$a_2(v)=1/\lambda$, which leads to the Fokker--Planck equation at small
$\delta$, 
\begin{subequations}
\begin{equation}
\frac{\partial}{\partial t}\tilde{P}_{t}(x,v) = \frac{1}{2\lambda}\partial_v^2 
\tilde{P}_t(x,v) - \frac{\partial}{\partial x}\left\{v h(x, v)\tilde{P}_t(x, 
v)\right\}
\label{eq:cont active delta small 1}
\end{equation}
with the reflecting boundary condition
\begin{equation}
\frac{\partial}{\partial v}\tilde{P}_t(x,v)=0\qquad \textrm{for $v=\pm 1$}.
\end{equation}
\label{EQ:FokkerPlanckActiveSinglePart}
\end{subequations}

This Fokker--Planck equation is equivalent to the coupled Langevin equation
in \eq{eq:contiLangevin1} of the approximate dynamics. In consistence
with the analysis in \secti{SecApproximateDynamics}, this Fokker--Planck
equation depends on two parameters ($\lambda$ and $\Gamma_1$) given in
\eq{Eq:reducedParameters1}. However, as noted earlier, this is only
true in a certain range of parameters where the subleading terms are
negligible. In particular, the Fokker--Planck equation in 
\eq{EQ:FokkerPlanckActiveSinglePart} fails to describe the
passive limit ($\lambda = 0$), where the relevant parameter is different
\cite{KapferKrauth2015} from $\Gamma_1$.

To describe the passive limit, a diffusive scaling with $\delta$ is 
required
\begin{equation}
t=\Dt\, \delta^2, \qquad v=\frac{\epsilon}{\delta}, \qquad x=x\,.
\label{eq:diffusive scaling}
\end{equation}
Starting from \eq{eq:Master discrete} and following a similar procedure as for
the derivation of \eq{eq:cont active delta small 1}, leads to the well-known
Fokker--Planck equation for a passive particle in a potential
\begin{equation}
\frac{\partial \tilde{P}_{t}(x)}{\partial t} 
=\frac{1}{k_\text{B}T}\frac{\partial}{\partial 
x}[U'(x)\tilde{P}_t(x)]+\frac{\partial^2}{\partial x^2}[\tilde{P}_t(x)]\,.
\label{eq:FP passive single}
\end{equation}
A detailed derivation\cite{KlamserThesis} finds back that in the passive limit
the inverse-power-law potential is described by 
a single control parameter, as is well known\cite{KapferKrauth2015}.

In conclusion, the number of relevant parameters can be reduced when 
$\delta$ is small: the persistent limit has two 
relevant parameters ($\Gamma_1$ and $\lambda$) and the passive limit ($\lambda 
= 0$) has a single 
parameter $\Gamma_0$ (see \eq{eq:Gamma0}). However, $\Gamma_1$ does not 
converge to $\Gamma_0$ for $\lambda \to 0$. Therefore, the order of the 
limits $\delta \to 0$ and $\lambda \to 0$ cannot be exchanged.

\subsection{Multi-particle case\label{Multi-particle conti limit KMC}}
The discussion in \secti{S: exact analysis of MC} can be generalized to the 
multi-particle case. The single-particle Fokker--Planck
equation in \eq{eq:cont active delta small 1} generalizes to the $N$-particle
Fokker--Planck equation
\begin{equation}
\frac{\partial}{\partial t}P_{t}[\textbf{x,v}] = 
\frac{1}{2\lambda}\sum_i\frac{\partial^2}{  \partial v_i^2} P_{t}[\textbf{x,v}] 
- \sum_i\frac{\partial}{\partial x_i}\left\{v_i h_i(\textbf{x}, 
v_i)P_{t}[\textbf{x,v}]\right\}\,,
\label{eq:FP mutiparticle active}
\end{equation}
where particles interact via the inter-particle potential $U(\xvec)$, with
$ \xvec =\left\{ x_1,\cdots,x_N\right\}$ and
\begin{equation*}
h_i(\textbf{x}, v_i)=\min\left\{1,\exp(\Gamma_1 v_i F_i[\textbf{x}]) 
\right\}\,,
\end{equation*}
with the force on particle $i$
\begin{equation}
F_i[\textbf{x}]=-\frac{\partial U[\textbf{x}]}{\partial x_i}\,.
\end{equation}
The reflecting boundary condition in the velocity space is
\begin{equation}
\frac{\partial}{\partial v_i}P_{t}[\textbf{x,v}]=0\qquad \textrm{for $v_i=\pm 
1$.}
\end{equation}

The Fokker-Planck \eqref{eq:FP passive single} in the passive limit has a very 
similar multi-particle generalization.
\subsubsection*{Power-law interaction potential}
The numerical studies presented in this work are for the inverse-power-law
potential in \eq{eq:Potential}. The force on particle $i$ is
\begin{equation}
F_i[\textbf{x}]=n u_0\gamma^n\sum_{j\ne i}\frac{\text{sgn}(x_i-x_j)}{ 
|x_i-x_j|^{n+1}}\,.
\label{eq:ForcePoserLaw}
\end{equation}
For this specific choice, two dimensionless parameters
characterize the persistent many-particle behavior. These can be obtained from 
the mean inter-particle distance
\begin{equation}
d = \frac{L}{N} = \frac{\gamma}{\phi}\,, \qquad \text{for 1D}
\end{equation}
with $L$ being the system size or, equivalently, from the dimensionless 
density 
\begin{equation}
\phi = \frac{\gamma N}{L}\,, \qquad \text{for 1D.}
\end{equation}
Then, using the scaled coordinates
\begin{equation}
\tilde{t}=\frac{t}{d}, \qquad \tilde{v}(\tilde{t})=v(t), \qquad 
\tilde{x}(\tilde{t})=\frac{x(t)}{d},
\end{equation}
in \eq{eq:FP mutiparticle active}, we obtain the rescaled Fokker--Planck 
equation 
\begin{equation}
\frac{\partial}{\partial 
\tilde{t}}\tilde{P}_{\tilde{t}}[\tilde{\textbf{x}},\tilde{\mathbf{v}}] = 
\frac{1}{2\tilde{\lambda}}\sum_i\frac{\partial^2}{  \partial \tilde{v}_i^2} 
\tilde{P}_{\tilde{t}}[\tilde{\textbf{x}},\tilde{\textbf{v}}] - 
\sum_i\frac{\partial}{\partial \tilde{x}_i}\left\{\tilde{v}_i 
\tilde{h}_i(\tilde{\textbf{x}}, 
\tilde{\mathbf{v}}_i)\tilde{P}_{\tilde{t}}[\tilde{\textbf{x}},\tilde{\mathbf{v}}
]\right\}\,,
\label{eq:FP mutiparticle active d scaled}
\end{equation}
with
\begin{equation}
\tilde{h}_i(\tilde{\xvec}, \tilde{v}_i)=\min\left\{1,\exp\left(\Gamma 
\tilde{v}_i 
\sum_{j\ne i}\frac{\text{sgn}(\tilde{x}_i-\tilde{x}_j)}{( 
\tilde{x}_i-\tilde{x}_j)^{n+1}}\right)\right\}\,,
\end{equation}
and
\begin{equation*}
 \tilde{\lambda}=\frac{\lambda}{d}\,,\quad \text{and} \quad \Gamma=\frac{n u_0
 \gamma^n \delta}{k_\text{B}Td^{n+1}}\,.
\label{eq: active gamma}
\end{equation*}
These two parameters ($\lambdatilde$ and $\Gamma$) govern the scaled 
many-particle probability distribution. 

\begin{figure}
\centering
\includegraphics[width = 0.4 
\textwidth]{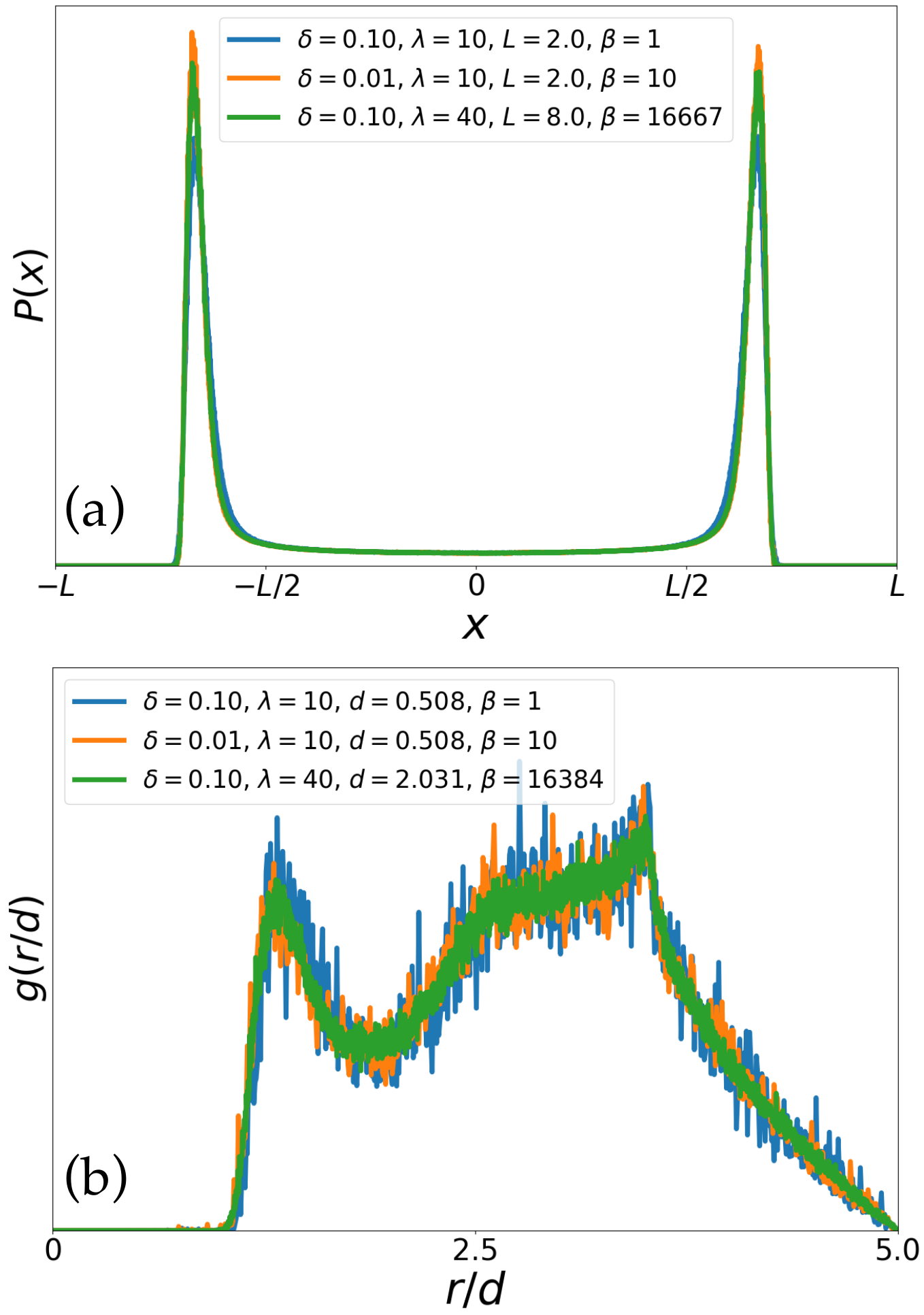}
\hfill %
\caption{Numerical verification of the reduced set of relevant parameters in the
active limit (finite $\lambda$).
\subcap{a}: 1D single-particle case. Histogram of the 
particle position in a confining potential $U(x) = u_0 \gamma^6 ((L+x)^{-6} + 
(L-x)^{-6})$. 
Data for different $\delta$, $\lambda$ and $\beta$, but constant 
$\lambda/L = 2.5$ and 
$\beta u_0 \delta \gamma^6/L^7 = 6.209 \times 10^{-6}$.  
\subcap{b}:~Two-dimensional many-particle case. Pair-correlation function 
$g(r)$ for $N = 
16$ and $n = 6$. Data for different $\delta$, $\lambda$ and $\beta$, at
constant 
 $\lambda/d = 11.1$ and $\beta u_0 \delta \gamma^6/d^7
= 0.209$. (All data for $\gamma = 1$, $u_0 = 1$.)
}
	\label{F_ScalingActiveScaling}
\end{figure}

It is useful to express these parameters in 
terms $\phi$, which leads to
\begin{equation}
\tilde{\lambda}=\frac{\delta^3}{\sigma^2\gamma}\phi\,\,\quad \text{and}\quad 
\Gamma=\frac{u_0}{k_\text{B}T}\frac{n\delta}{\gamma}\phi^{n+1}\,.
\label{EQ:scale1}
\end{equation}
The generalization to higher dimensions is straightforward. For example, in 
2D, where $\phi = N\gamma^2/V$ and $d = \sqrt{V/(N\pi)} = \gamma/\sqrt{\pi 
\phi}$, 
the two relevant parameters are
\begin{equation}
\tilde{\lambda}=\frac{\delta^3}{\sigma^2\gamma}\sqrt{\pi \phi}\,\,\quad  
\text{and}\quad \Gamma=\frac{u_0}{k_\text{B}T}\frac{n\delta}{\gamma}(\pi 
\phi)^{(n+1)/2}\,.
\label{EQ:scale2222}
\end{equation} 
In this form, the dimensionality primarily enters into the $\phi$-dependence of 
these two parameters (see \eq{EQ:scale1}).
The validity of these reduced parameter sets is confirmed by numerical 
simulations (see \fig{F_ScalingActiveScaling}).

As discussed earlier in the single-particle case, this parameter reduction
does not extend to the passive regime. In this case a diffusive scaling is
required
\begin{equation}
\tilde{t}=\frac{t}{d^2}, \qquad \tilde{x}(\tilde{t})=\frac{x(t)}{d}\,,
\end{equation}
in terms of which the Fokker--Planck equation becomes\cite{KlamserThesis}
\begin{equation}
 \frac{\partial}{\partial \tilde{t}}\tilde{P}_{\tilde{t}}[\tilde{\textbf{x}}] = 
-\frac{1}{6}\Gamma_0\sum_i\frac{\partial}{\partial 
\tilde{x}_i}\left\{\sum_{j\ne 
i}\frac{1}{( \tilde{x}_i-\tilde{x}_j)^7} 
P_{\tilde{t}}[\tilde{\textbf{x}}]\right\}+\frac{1}{6}\sum_i\frac{\partial^2}{  
\partial \tilde{x}_i^2} P_{\tilde{t}}[\tilde{\textbf{x}}] \,,
\label{eq:FP mutiparticle passive double scaled}
\end{equation}
with 
\begin{equation}
\Gamma_0= \begin{cases}
                        \frac{n u_0 }{k_\text{B} T} \phi^n \,, \text{ in 1D} \\
                        \frac{nu_0}{k_\text{B} T}(\pi \phi)^{n/2}\,, \text{ in 
2D} \,.
                    \end{cases}
\label{eq:Gamma0}
\end{equation}
Therefore, the probability only depends\cite{KapferKrauth2015} on 
$\Gamma_0$, which differs from $\Gamma$.

In summary, the small-$\delta$ limit of the systems described in this work 
is described by 
two relevant parameters for the active regime and a single relevant parameter 
for the passive regime. There is no smooth transition from one regime to the 
other when $\delta$ is small. This is because the limits $\delta \to 0$ and 
$\lambda \to 0$ are not exchangeable. 
Therefore, a phase diagram covering both passive and active is not possible 
on a two parameter space. A complete phase diagram is three dimensional on a 
rescaled parameter space $\delta/d$, $\lambda/d$ and $\Gamma_0$ in 
\eq{eq:Gamma0}. Expressed in terms of $\phi$, they yield
\begin{align*}
\frac{\delta}{\gamma}\phi\,, \frac{\lambda}{\gamma}\phi\,, &\text{  and  } 
\Gamma_0\,, \quad \text{for 1D}\\
\frac{\delta}{\gamma}\sqrt{\pi\phi}\,, \frac{\lambda}{\gamma}\sqrt{\pi\phi}\,, 
&\text{  and  } \Gamma_0\,, \quad \text{for 2D.}
\end{align*}

\subsubsection*{Continuous-time description}
In the analysis presented so far, the Fokker--Planck description was
obtained in the small-$\delta$ limit. However, it is 
possible\cite{KlamserThesis}
to define a continuous-time description of the MC dynamics for arbitrary
values of $\delta$. This can be obtained using the Kramers--Moyal
expansion\cite{StochasticProcessesVanKampen} of the Master equation
\eq{eq:Master discrete}, which leads to the Fokker--Planck equation
\begin{align}
\partial_t P_t(x,\epsilon) = &\frac{\sigma^2}{2} 
\frac{\partial^2}{\partial\epsilon^2}P_t(x,\epsilon) +
f(x-\epsilon,\epsilon)P_t(x-\epsilon,\epsilon)\cr
&-f(x,\epsilon)P_t(x,\epsilon)\,,
\label{eq:Fokker--Planck general}
\end{align}
with the reflecting boundary at 
$\epsilon = \pm\delta$ as zero-current condition 
\begin{equation}
\frac{\partial P_t(x,\epsilon)}{\partial\epsilon}\bigg 
\vert_{\epsilon=\pm\delta} = 0\,.
\label{eq:Fokker--Planck general_boundary}
\end{equation}
Taking the small-$\delta$ limit of
\eq{eq:Fokker--Planck general}, we recover the scaling of \eq{EQ:scale2222} 
that was obtained from the discrete-time Master equation \eq{eq:Master 
discrete}.

\subsection{Active random acceleration process}
An intriguing aspect of the dimensional reduction of the phase 
space is the singular nature of the passive case on a two-parameter plane for 
$\delta \to 0$. This singular behavior of the rescaled parameters is 
not due to the discrete nature of the MC dynamics as it also appears 
in the continuous-time description discussed above. For the kinetic MC, 
the singularity appears in the small-$\delta$ limit.  However, it can also 
appear in other stochastic models of active matter, even without taking a 
vanishing limit of a parameter similar to $\delta$. To illustrate this point, 
we define a continuous model which closely resembles the kinetic MC dynamics. 
This model, which we refer as the active random acceleration 
process\cite{RandomAccelerationProcess} is defined for continuous time by a 
coupled Langevin equation
\begin{subequations}
\begin{align}
\dot{\epsilon}_t &= r_t + R\left(\frac{\epsilon_t}{\delta}\right)\,,\label{eq: 
eps Langevin ARAP}\\
\dot{x}_t &= v_0\epsilon_t + \frac{1}{k_\text{B}T} \, F(x_t) + s_t\,, 
\label{eq: 
x Langevin ARAP}
\end{align}
\label{Eq:ARAP}
\end{subequations}
where $x_t$ is the position of the self-propelled particle at time $t$. The two 
noise terms, $r_t$ and $s_t$, are Gaussian white noises with zero mean and 
covariance $\langle r_t r_{t'}\rangle = \sigma^2 \delta(t-t')$ and 
$\langle 
s_t s_{t'}\rangle = 2D \delta(t-t')$, respectively. The term $R$ denotes 
the reflecting boundary condition at $\pm \delta$. The parameter $v_0$ 
characterizes the strength of the self-propulsion\footnote{Putting 
$1/k_\text{B}T$ in the force term of \eq{eq: x Langevin ARAP} is motivated by 
the aim to compare with the kinetic MC results.}.

Similar to the kinetic MC dynamics, in the scaled coordinates $t\to t d / 
\delta$, $\epsilon \to v \delta$, $x \to x d$ the dynamics \eq{Eq:ARAP} is 
described by the Fokker--Planck equation 
\begin{align}
\label{eq: rescaled ARAP - dimensionless FP}
\frac{\partial}{\partial t}P_t(x,v) = \frac{d}{2\lambda} 
\frac{\partial^2}{\partial v^2} P_t(x,v) + \frac{D}{\delta d} 
\frac{\partial^2}{\partial x^2} P_t(x,v)\cr
- \frac{1}{\delta k_\text{B}T}  \frac{\partial}{\partial x} \left[ F(xd) 
P_t(x,v)\right] - v_0\frac{\partial}{\partial x} [v P_t(x,v)]\,.
\end{align}
Its multi-particle generalization is straightforward. For the inter-particle 
force  \eq{eq:ForcePoserLaw}, the steady-state probability distribution is 
controlled by three parameters
\begin{align}
A= \frac{D}{\delta d} \frac{\lambda}{d}\,,\, 
B = v_0 \frac{\lambda}{d}\,, \text{ and 
}\,C = \frac{n u_0 \gamma^n}{k_{\text{B}}T \delta d^{n+1}} 
\frac{\lambda}{d}\,.
\end{align}

The passive limit corresponds to $v_0 = 0$, where the $x$ and $v$ coordinates 
decouple and the probability $P(x,v) = P(x)P(v)$. In such case, we see that the 
$P(x)$ follows the usual equilibrium Fokker--Planck equation and the 
steady-state 
$P(x)$ is determined by a single control parameter $C/A$, which coincides with 
$\Gamma_0$ for $D = 1$. Then, importantly, $P(x)\propto 
\exp(- C 
u(x)/A)$, where $u(x)$ is the rescaled potential and therefore, setting $D = 0$ 
makes the probability distribution uniform and independent of the 
inter-particle 
interaction, which corresponds to an infinite temperature.

In the active limit, there are the three control parameters $A$, $B$, and $C$. 
The only way to reduce the number of parameters, while keeping the 
particles 
active and interacting, is by setting $D = 0$ and thereby $A = 0$. However, 
this 
would make the finite-temperature passive limit inaccessible, as we discussed 
above. This shows that the singular behavior of the MC dynamics, 
when trying to reduce the number of control parameters to two also happens in 
this active random acceleration process. A similar singular behavior is present 
for other models. We have checked\cite{KlamserThesis} that the same statement 
applies for the active Ornstein--Uhlenbeck process.

From this analysis, we can conclude that in these classes of active dynamics, 
at least three independent relevant parameters are required to describe 
the full phase diagram including the passive regime.

\section{Conclusions\label{Conclusions}}

In this work, we presented a kinetic-Monte Carlo perspective on 
two-dimensional active 
matter. Within this approach, we established (in extension of our earlier work
\cite{KlKaKr2018}) the presence of the liquid, hexatic, and solid 
active-matter phases from their constituent decay laws of positional and 
orientational order (see 
\tab{tab:PhaseCharacteristics}). We also ascertained continuity of the 
active-matter phases in the passive limit and recovered the phases of the  
corresponding equilibrium system. 
We have not tested very soft 
repulsive inverse-power-law potentials ($n < 6$ in \eq{eq:Potential}), but 
expect on the 
basis of our findings that for sufficiently steep repulsive 
inverse-power-law potentials the two-step melting 
behavior of the equilibrium system is maintained up to high activities, and 
possibly up to infinite persistence lengths. 
The stability of the intermediate hexatic phase in the high-activity 
regime---far above the linear-response regime---is 
intriguing. It may be due to an underlying symmetry that is yet to be 
understood. We have not addressed here 
the nature of the melting transitions in the active region, that is, the 
question of whether the liquid--hexatic phase transition is  of 
first order or of Kosterlitz--Thouless type\cite{KT1,KosterlitzThouless1973} 
(as in the equilibrium system \cite{KapferKrauth2015}), and whether 
this theory continues to apply at all to the hexatic--solid transition in 
active systems.

Besides the melting phase transitions, we have established the existence of 
the MIPS phase and identified it as 
a coexistence between a liquid and a gas: two phases with
exponential decays of both correlation 
functions, but 
with typically very  different densities. The 
coexistence region was identified for a wide range of interaction potentials
including the hard-disk limit.
It remains disjoint from the melting lines for all 
values of $n$ (possibly excluding very soft potentials).
An analogy of the MIPS with the liquid--gas coexistence in equilibrium 
indicates the existence\cite{SpeckCritical} of 
a critical point, although we have not studied it in detail.

We also discussed the dimensionality of 
the phase diagram. We showed that the qualitative 
phase behavior is robust against changes of the maximum step size $\delta$.
The steady state reached by the kinetic MC dynamics is fully described
by the density, the persistence length, and the maximum step size $\delta$, 
although 
a two-dimensional scaling describes the MIPS phase and the melting 
transitions at high density $\phi$ for small $\delta$. We 
argued in this direction using 
the stochastic description of the kinetic MC dynamics, and its formulation in 
terms of a Langevin dynamics. The common scaling with $\delta$ of the 
melting transitions and of the MIPS breaks down in the vicinity of the 
equilibrium phase 
transition point. This is needed to obtain the one-parameter scaling (for an 
inverse-power-law potential) in the passive limit. 

The detailed phase diagrams found in our present  work once more illustrate the 
rich collective properties in non-equilibrium physics, and the current 
limits of our understanding of these models and, more generally, of the physics 
of active matter.

\begin{acknowledgments}

We thank H. Löwen and L. Berthier for helpful discussions. W.K. acknowledges 
support from the Alexander von Humboldt Foundation.
\end{acknowledgments}

\appendix
\section{Passive Fokker--Planck equation under diffusive 
scaling\label{S:PassiveDiffusive}}
In the passive limit, it follows from \eqref{DisplacementDistribution} that 
$g(\epsilon,\epsilon')=1/ (2 \delta)$. In this case, the Master 
equation  (of 
\eq{eq:Master discrete}) can be recast by using the definition 
$P_\Dt(x)=\int_{-\delta}^\delta d\epsilon P_\Dt(x,\epsilon)$ and integrating 
over $\epsilon'$ leads to
\begin{align}
&P_{\Dt+1}(x) =P_{\Dt}(x)+ \frac{1}{2\delta} \int_{-\delta}^\delta d\epsilon \cr
& \left\{f(x-\epsilon,\epsilon)P_\Dt(x-\epsilon)-f(x,\epsilon) 
P_\Dt(x)\right\}\,.
\label{eq: so}
\end{align}
Applying the diffusive 
scaling in \eq{eq:diffusive scaling}, it follows
\begin{align*}
&\tilde{P}_{t+\delta^2}(x) =\tilde{P}_{t}(x)+\frac{1}{2}\int_{-1}^{1} dv \cr
&\left\{f(x-\delta v,\delta v)\tilde{P}_t(x-\delta v)-f(x,\delta v) 
\tilde{P}_t(x)\right\}\,,
\end{align*}
with $\tilde{P}_t(x) = P_\Dt(x)$.
Expanding in powers of small $\delta$ leads to
\begin{equation}
\tilde{P}_{t+\delta^2}(x) 
=\tilde{P}_{t}(x)-\frac{\delta}{2}\frac{d}{dx}[A(x)\tilde{P}_t(x)]+\frac{
\delta^2}{4}\frac{d^2}{dx^2}[B(x)\tilde{P}_t(x)]+\cdots\,,
\label{eq:intermediateP}
\end{equation}
with the definitions
\begin{equation}
A(x)=\int_{-1}^{1}dv\, v\, f(x,\delta v)\,\text{, and }\, 
B(x)=\int_{-1}^{1}dv\, v^2\, f(x,\delta v)\,.
\label{eq:A and B}
\end{equation}
The expression for $f(x,\epsilon)$ in \eq{EQ_ApproxDynam} can be rewritten as 
\begin{align*}
f(y,\epsilon) 
=1-\Theta\bigg(\frac{U(y+\epsilon)-U(y)}{k_\text{B}T}\bigg)\cr
\bigg\{
1-\exp\left(-\frac{U(y+\epsilon)-U(y)}{k_\text{B}T}\right)\bigg\}\,,
\end{align*}
with $\Theta(x)$ the Heaviside function.  Using this expression in
$A(x)$ in \eq{eq:A and B} and expanding in terms of small $\delta$, it follows
\begin{align*}
A(x)= \beta \delta  F(x) \int_{-1}^{1}dv\, 
v^2\,\Theta\bigg(- 
v\, 
F(x)\bigg)+\cdots\,,
\end{align*}
 with the force $F(x) = - \tfrac{\partial}{\partial x} U (x)$. The two cases 
$F(x)>0$ and $F(x)<0$ lead to the same result for the integral (namely $1/ 3$), 
thus giving
\begin{equation}
A(x)= \beta \frac{\delta \, F(x)}{3}+\cdots\,.
\label{eq:A}
\end{equation}
In the same way it follows for $B(x)$ in \eq{eq:A and B} (the term with
$\Theta(..)$ contributes to order $\delta$ and is thus neglected)
\begin{equation}
B(x)=\frac{2}{3}+\mathcal{O}(\delta)
\label{eq:B}
\end{equation}
Using this expressions of \eqtwo{eq:A}{eq:B} in
\eq{eq:intermediateP} results in the well-known Fokker--Planck equation
for a passive particle in a potential 
\begin{equation} 
\frac{\partial
\tilde{P}_{t}(x)}{\partial t} =-\frac{1}{k_\text{B}T}\frac{\partial}{\partial
x}[F(x)\tilde{P}_t(x)]+\frac{\partial^2}{\partial x^2}[\tilde{P}_t(x)]\,.
\end{equation}
This corresponds to the standard Langevin description of a passive particle
\begin{equation}
\dot{x}(t)=\beta F(x)+\eta(t)\qquad \langle \eta(t)\eta(t')\rangle=2 
\delta(t-t').
\label{eq: passive langevin equation KMC}
\end{equation}
The steady-state probability $P(x)\sim \expb{-\beta U(x)}$, thus 
the single relevant parameter is $\beta U(x)$.

If the ballistic scaling of \eq{Eq:BallisticScaling} was used for the passive 
case, it would result in
\begin{equation}
\frac{\partial P_{t}(x)}{\partial t} 
=\frac{\Gamma_1}{6}\frac{\partial}{\partial x}[F(x)P_t(x)]\,,
\label{eq: Passive KMC zero Temp}
\end{equation}
which clearly does not capture the correct physics.

\bibliography{References,General}

\end{document}